\documentclass[journal=jacsat,manuscript=article]{achemso}

\usepackage{chemformula} 
\usepackage[T1]{fontenc} 
\usepackage{empheq}

\author{Cinzia Di Giorgio}
\affiliation{Department of Physics E.R. Caianiello, University of Salerno, Fisciano, 84084 Italy}
\alsoaffiliation{Laboratoire de Physique des Solides, CNRS, Université Paris-Saclay, Orsay, 91405 France}
\altaffiliation{Current address: Materials Foundry Institute, National Research Council (CNR-IOM), Trieste, 34149 Italy}

\email{digiorgio@iom.cnr.it}

\author{Elena Blundo}
\affiliation{Physics Department, Sapienza University of Rome, Rome, 00185 Italy}

\author{Julien Basset}
\affiliation{Laboratoire de Physique des Solides, CNRS, Université Paris-Saclay, Orsay, 91405 France}

\author{Giorgio Pettinari}
\affiliation{Institute for Photonics and Nanotechnologies, National Research Council (CNR-IFN), Rome, 00133 Italy}

\author{Marco Felici}
\affiliation{Physics Department, Sapienza University of Rome, Rome, 00185 Italy}

\author{Charis H.L. Quay}
\affiliation{Laboratoire de Physique des Solides, CNRS, Université Paris-Saclay, Orsay, 91405 France}

\author{Stanislas Rohart}
\affiliation{Laboratoire de Physique des Solides, CNRS, Université Paris-Saclay, Orsay, 91405 France}

\author{Antonio Polimeni}
\affiliation{Physics Department, Sapienza University of Rome, Rome, 00185 Italy}

\author{Fabrizio Bobba}
\affiliation{Department of Physics E.R. Caianiello, University of Salerno, Fisciano, 84084 Italy}
\alsoaffiliation{SuPerconducting and other INnovative materials and devices institute, National Research Council (CNR-SPIN), Fisciano, Italy}

\author{Marco Aprili}
\affiliation{Laboratoire de Physique des Solides, CNRS, Université Paris-Saclay, Orsay, 91405 France}

\title[]
  {Imaging the Quantum Capacitance of Strained MoS$_2$ Monolayers by Electrostatic Force Microscopy}

\keywords{2D materials, MoS$_2$, Strain, Quantum Capacitance, Electrostatic Force Microscopy}

\begin{document}







\begin{abstract}
  We implemented radio frequency-assisted electrostatic force microscopy (RF-EFM) to investigate the electric field response of biaxially strained molybdenum disulfide (MoS$_2$) monolayers (MLs) in the form of mesoscopic bubbles, produced via hydrogen (H)-ion irradiation of the bulk crystal. MoS$_2$ ML, a semiconducting transition metal dichalcogenide, has recently attracted significant attention due to its promising opto-electronic properties, further tunable by strain. Here, we take advantage of the RF excitation to distinguish the intrinsic quantum capacitance of the strained ML from that due to atomic scale defects, presumably sulfur vacancies or H-passivated sulfur vacancies. In fact, at frequencies $f_{RF}$ larger than the inverse defect trapping time, the defect contribution to the total capacitance and to transport is negligible. Using RF-EFM at $f_{RF}=300$ MHz, we visualize simultaneously the bubble topography and its quantum capacitance. Our finite-frequency capacitance imaging technique is non-invasive and nanoscale, and can contribute to the investigation of time and spatial-dependent phenomena, such as the electron compressibility in quantum materials, which are difficult to measure by other methods.\\

  \textit{Keywords}: 2D materials, MoS$_2$, Strain, Quantum Capacitance, Electrostatic Force Microscopy
\end{abstract}

\section{Introduction}
Two-dimensional (2D) transition metal dichalcogenides (TMD) have recently attracted significant attention due to their promising opto-electronic and mechanical properties and ultrathin nature \cite{Manzeli2017}. As in graphene, the weak van-der Waals (vdW) coupling between adjacent layers enables scalability from thick TMDs to mono- or few-layer crystals \cite{Manzeli2017}. This reduction in dimensionality, from bulk to monolayer (ML), has been shown to affect the material's band structure, driving bandgap transitions, and modifying the bandgap width \cite{Mak2010, Jin2013, Komsa2012, Trainer2017}. TMDs at the atomic scale are of interest for device scaling \cite{Radisavljevic2011, Baugher2013, Kim2012, Desai2016, Jun2014, Ghatak14}. In particular, ML to few-layer molybdenum disulfide (MoS$_2$)-based transistors have demonstrated impressive device performance, characterized by high mobility, low sub-threshold slope, and excellent on/off current ratios \cite{Radisavljevic2011, Ghatak14}. Moreover, as the band structure of TMD MLs is strongly affected by mechanical deformations\cite{Naumis2017, Roldán2015, Guinea2009, Blundo2021}, the performance of TMD-based devices can be further tuned by strain \cite{Datye2022}. Given their high in-plane stiffness and low flexural rigidity \cite{Akiwande2017}, individual atomic sheets of TMDs can sustain much larger mechanical strains than conventional semiconductors \cite{Changgu2088}, allowing wider and finer tuning of electronic and photonic properties \cite{DiGiorgioREV}. However, their atomic scale thickness makes their intrinsic properties, and related device performance, very sensitive to structural defects, such as atomic vacancies, dislocations, and grain boundaries \cite{Hu2018, Yin2019}, which are ubiquitous and sometimes further induced by strain production techniques. Some defects can behave as charge traps, introducing mid-gap and/or band edge states \cite{Qiu2013, Hong2015, Liu2013}. A Shockley-Read-Hall (SRH)-like exchange of electrons or holes, can occur between the defect state and the majority carrier band, affecting the capacitive response of the material to an external electric field \cite{Brammertz2008, Zhu2014}. This a priori uncontrollable \textit{defect-state capacitance} adds to the intrinsic \textit{quantum capacitance} ($c_Q$), which accounts for the partial screening of electric fields in 2D materials, due to their low density of states (DOS) \cite{Luryi1988, Bera2019, Ma2015, Fang2018}. 

In this article, we investigate the quantum capacitance of biaxially strained MoS$_2$ MLs in the form of 'bubbles': low-energy hydrogen (H)-ion irradiation induces (sub-)micrometric protrusions where a ML detaches from the bulk \cite{Tedeschi2019, He2019, DiGiorgio2020, DiGiorgio2021, Blundo2020}. These bubbles are filled with highly pressurized H$_2$ gas \cite{DiGiorgio2020, DiGiorgio2021, Blundo2020, Blundo2021, Blundo2021PRL, Blundo2022}. Low-energy H-plasma treatments induce sulfur vacancies at the topmost interface of the MoS$_2$ ML, with the Mo and S atoms in both the underneath planes left untouched \cite{Ghorbani2017, Lee2021}. In addition, H-intercalation can also partially passivate sulfur vacancies in MoS$_2$ ML \cite{Pierucci2017}. Furthermore, the H adsorption free energy of a MoS$_2$ surface is affected by the density of sulfur vacancies, as well as by strain, decreasing as these increase \cite{Li2016}. The presence and/or passivation of atomic-scale defects is thus expected to contribute to the electrostatic properties of the MoS$_2$ bubbles.

Here, we study the electrostatic properties of bulk MoS$_2$ and MoS$_2$ bubbles via radio frequency-assisted electrostatic force microscopy (RF-EFM). EFM is a on non-contact atomic force microscopy (AFM)-based technique. 
We used an approach similar to the so-called «high–low Kelvin probe force microscopy/spectroscopy (HL-KPFM)», recently introduced
by Izumi R. et al.\cite{Izumi2023, Sugawara2020}. In fact, by studying the long-range electrostatic interaction between MoS$_2$ and a conducting Pt tip, biased with (DC+RF) voltage, we distinguish the capacitive contribution due to the finite DOS (quantum capacitance) from that due to defects. Indeed, we demonstrate that at low frequency (sub-MHz range), the capacitance of the MoS$_2$ ML comes from the defect states. At higher frequencies (larger than the inverse defect trapping time), the intrinsic quantum capacitance dominates. Differently from HL-KPFM\cite{Izumi2023, Sugawara2020}, RF-EFM allows to span over a continuous frequency range of the AC bias voltage, and, unlike existing macroscopic capacitance-voltage measurements \cite{Brammertz2008, Fang2018}, which average over entire devices \cite{Giannazzo2009, Tino2013, Li2022}, this method allows us to directly image $c_Q$, over the bubble surface with a sub-micrometric spatial resolution. Finally, from the quantum capacitance we derive the Fermi energy $E_F$, which gives us the local charge carrier density. Through our analysis, we  elucidate the correlation between defects, band structure and carrier density, over the ML-thick membrane. 

The use of RF-EFM for finite-frequency capacitance mapping can be useful in the investigation of other phenomena as well. The nanoscale electron compressibility in quantum materials, the evolution of conductance domains at the metal-insulator transition \cite{Liu2020}, the defect states of perovskites-based harvesting systems \cite{Futscher2020, Yun2022}, the frequency dependence of the relative dielectric constant in polymeric insulators (commonly implemented in metal-insulator-semiconductor structures \cite{Estrada2013}), etc. Moreover, as EFM is nondestructive both for tip and  sample surface, it is is robust and reproducible.

\section{Results and discussion}

\begin{figure}
    \centering
    \includegraphics[width=16cm]{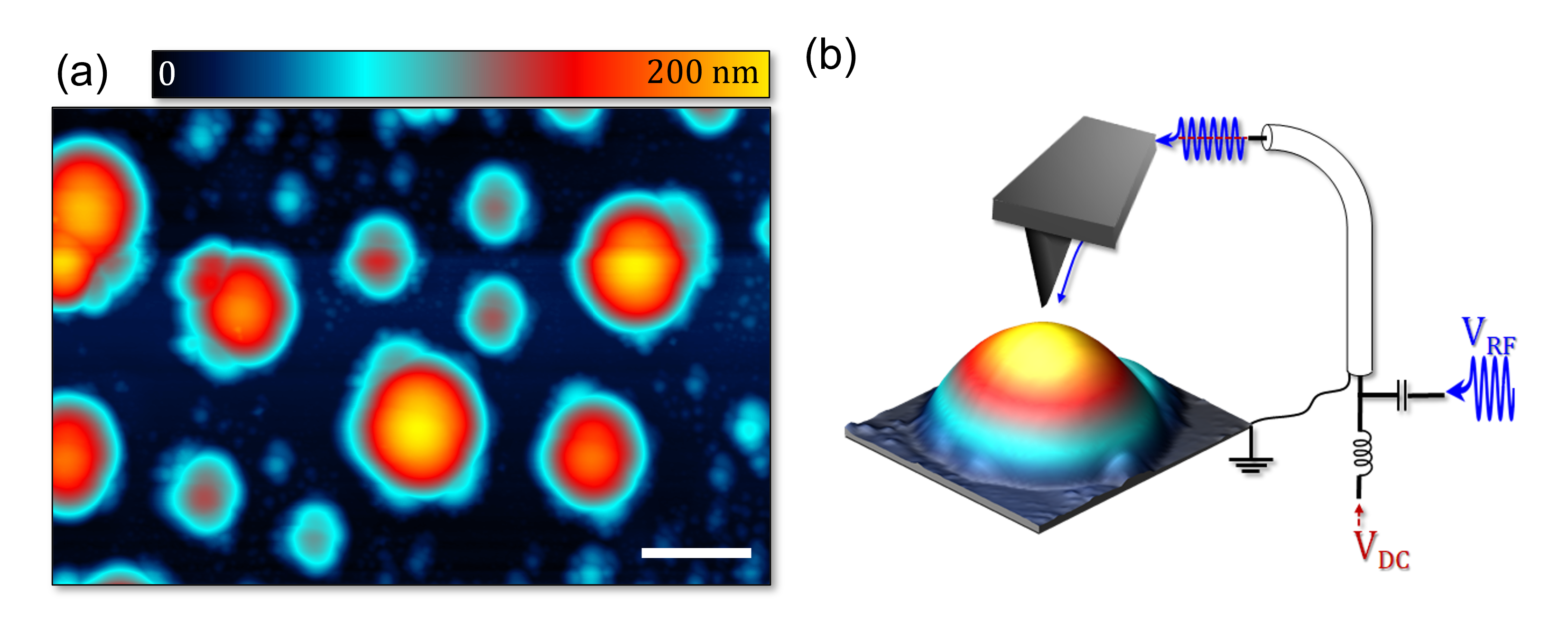}
    \caption{(a) AFM topography, $\mathrm{13\ \mu m \times 9.2\ \mu m}$ in lateral size, of an irradiated MoS$_2$ bulk crystal. Scale bar is $ \mathrm{2\ \mu m}$. (b) Scheme of RF-EFM.}
    \label{fig:asimmetry}
\end{figure}

Figure \ref{fig:asimmetry}(a) shows a representative ($\mathrm{13\ \mu m \times 9.2\ \mu m}$) AFM topography of the surface of an irradiated MoS$_2$ bulk crystal. Numerous H$_2$-filled bubbles can be seen, which have an universal (\textit{i.e.}, not dependent on their size) ratio of maximum height $h_{max}$ to radius $R$, which determines their strain profile \cite{Khestanova2016} (see Supporting Information 1, and Refs. \cite{Tedeschi2019, DiGiorgio2020, DiGiorgio2021, Blundo2020, Blundo2021, Blundo2021PRL}). The bubbles are separated by flat regions where the bulk crystal has remained intact after irradiation. Thus, both bulk MoS$_2$ and bubbles can be studied in the same sample. As shown in figure \ref{fig:asimmetry}(b), RF-EFM is a non-contact imaging technique based on the electrostatic interaction between the sample and a conducting AFM probe, biased with a (DC+RF) voltage, $V_{DC}+V_{RF}cos(2 \pi f_{RF}t)$, $f_{RF}$ being the frequency of the RF bias voltage. Note that the RF component can be turned on and off depending on measurements and, when on, $f_{RF}>0.1$  MHz, while the free resonance frequency of the used cantilevers, $f_0$, is $5 \div 11$ kHz. Therefore $f_{RF}>>f_0$ which allows us to separate the dynamics of carriers in MoS$_2$ from that of the cantilever. The cantilever oscillation amplitude is typically 10 nm and the tip-sample distance is 100 nm or larger, corresponding to negligible electron tunneling between the tip and the sample. To investigate the MoS$_2$ electric field response, we determine the tip-sample electrostatic force $F_{el}$ by measuring the frequency shift, $df=f-f_0$, of the cantilever as a function of the bias voltage $V=V_{DC}+V_{RF}cos(2 \pi f_{RF}t)-V_0$, with $-eV_0=W_T-W_S$ being the mismatch between sample and tip work functions, $W_S$ and $W_T$ respectively \cite{Sadewasser2018, Melitz2011}. $df$ is related to $F_{el}$ through $df=-\frac{f_0}{2k}\frac{\partial F_{el}}{\partial z}=-\frac{f_0}{4k}\frac{{\partial }^2}{{\partial z}^2}\left(c_T{V_T}^2\right)$. Here, $c_T$ and $V_T$ are the tip-surface capacitance and voltage drop, respectively, and $k$ is the cantilever spring constant. In general, $V_T=V-V_S=\frac{c_S}{c_S+c_T}V$, with $V_S$ and $c_S$ being the sample to ground voltage drop (also called surface potential) and capacitance, respectively.  As reported in Ref. \cite{Cowie2022} , $V_S$ is the voltage induced band bending. In the metallic limit (i.e. good screening), $c_S >> c_T$, $V_T=V$, and $df=-\frac{f_0}{4k}V^2\frac{{\partial}^2 c_T}{{\partial z}^2}$. If, instead, the screening is poor, $c_S$ approaches $c_T$, $V_T<V$, and $df=-\frac{f_0}{4k}V^2\frac{{\partial}^2}{{\partial z}^2}\left[c_T\left(\frac{c_S}{c_S+c_T}\right)^2\right]$. In general $c_S$ is a function of $V_S$ and, in intrinsic 2D semiconductors, it reduces to the quantum capacitance : $c_Q (V_S)=e^2g_{2D}\left[1+\frac{exp(E_g/{2k_BT})}{2cosh(eV_S/{k_BT})}\right]^{-1}$ \cite{Ma2015, Fang2018}. In this case $c_Q$  is electron-hole symmetric, hence it is an even function of $V_S$. 
 
\begin{figure}
    \centering
    \includegraphics[width=9cm]{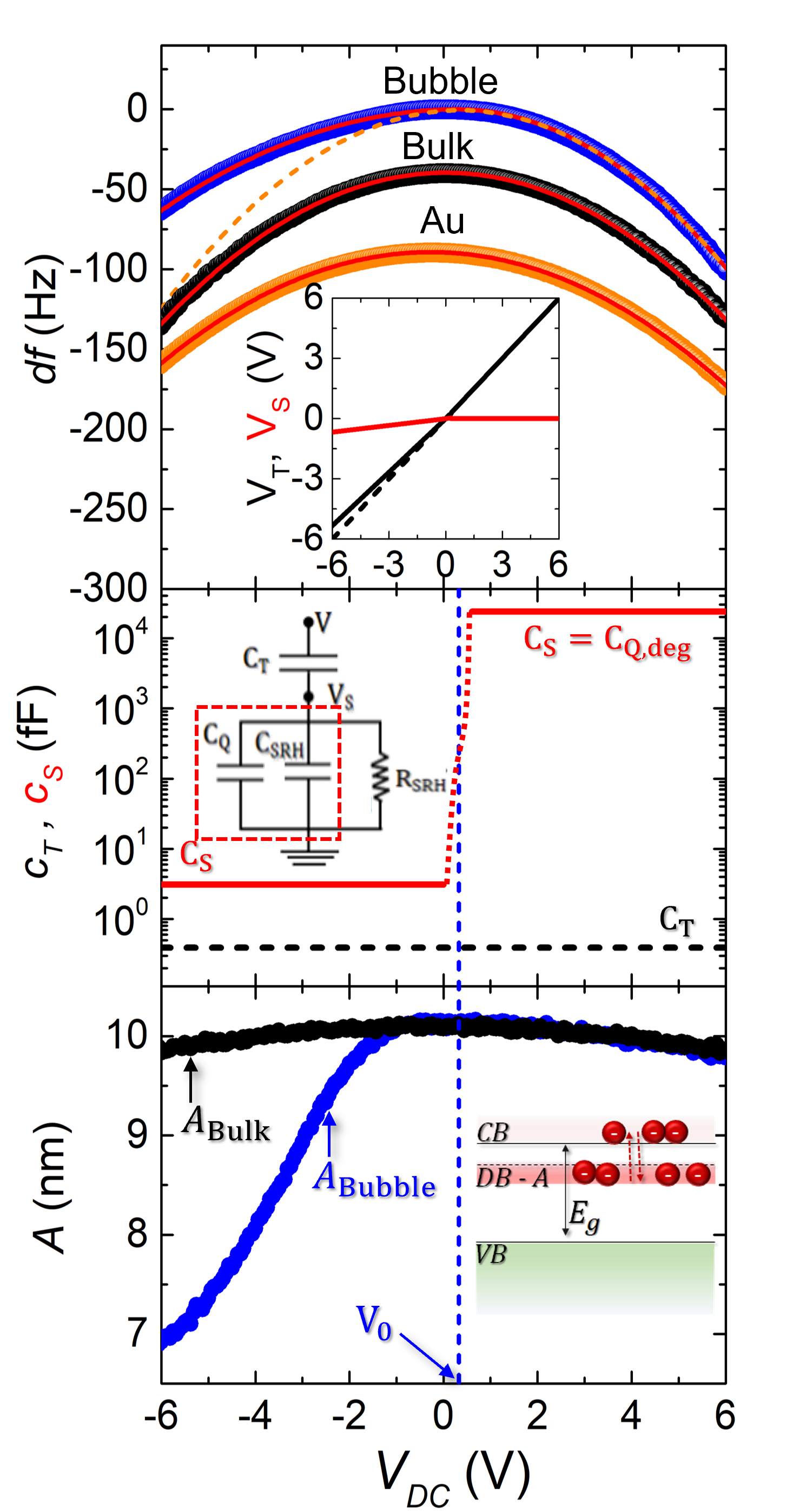}
    \caption{Top: (Main) $df(V_{DC})$ spectroscopy on the top-most location of a MoS$_2$ bubble (blu dots), MoS$_2$ bulk (black dots), and Au (orange dots), at tip-surface separation of 100, 125 and 135 nm, respectively. The red solid lines are fits to $df=-\frac{f_0}{4k}V^2\frac{{\partial}^2 c_T}{{\partial z}^2}$ for MoS$_2$ bulk and Au; and to $df=-\frac{f_0}{4k}V^2\frac{{\partial}^2}{{\partial z}^2}\left(\frac{c_Tc_S}{c_S+c_T}\right)$ for MoS$_2$ bubble. The orange dashed line is a representative fit to $df=-\frac{f_0}{4k}V^2\frac{{\partial}^2 c_T}{{\partial z}^2}$. (Inset) $V_T$ (black solid line) and $V_S$ (red solid line) vs $V_{DC}$. Black dashed line is used to show the deviation from $V_T=V_{DC}$. Middle: $c_T$ (black dashed line) and MoS$_2$ bubble $c_S$ (red plot) vs $V_{DC}$ (Main). Equivalent circuital configuration of the EFM experiment on MoS$_2$ bubble (Inset) (The same equivalent circuit is proposed by Ref. \cite{Izumi2023} in HL-KPFM). Bottom: cantilever oscillation amplitude $A$ vs $V_{DC}$ (Main). Band diagram of the MoS$_2$ bubble (Inset).}
    \label{fig:parabola}
\end{figure}

The top panel of figure \ref{fig:parabola} shows the $df(V_{DC})$ spectra of an irradiated MoS$_2$ sample, taken at a tip-surface separation of $z=100 \div 125$  nm, in two different locations: the bulk (black dots) and a bubble (blue dots). Additionally, the $df(V_{DC})$ spectrum of Au, taken at similar $z$, is reported (orange dots) for the sake of comparison with a perfect metal. The curve maxima are shifted from $V_{DC}=0$ by few tens of mV ($V_0=0.04, 0.3$ mV for MoS$_2$ bulk and ML and $V_0=-0.25$ mV for Au - see Supporting Information 2 for $V_0$ mapping over the bubble surface). While MoS$_2$ bulk and Au exhibit the expected quadratic voltage dependence of $df$ on $V_{DC}$ (parabolic fits in red), the $df(V_{DC})$ of the bubble lacks mirror symmetry: the orange dashed line, being the $V_{DC}^2$ fit of the data at positive $V_{DC}$, does not intercept the data on the negative $V_{DC}$ branch. This deviation from $V_{DC}^2$ is attributed to the breaking of the electron-hole symmetry in the band structure of MoS$_2$ MLs, due to the presence of defect states, as sketched in the inset of figure \ref{fig:parabola}-bottom panel. 

As suggested in Refs.\cite{Qiu2013,Hong2015,Liu2013}, atomic scale defects, such as sulfur vacancies, dislocations and grain boundaries, introduce a defect band (DB) close to the conduction band (CB). This results in an effective n-doping which moves the bottom of the CB closer to the Fermi level $E_F$ (or, vice versa, $E_F$ from the center of the bandgap, $E_g$, toward CB; in the following we will consider a fixed $E_F$ and movable CB minimum). The DB, previously observed in MoS$_2$ ML-based field effect transistors  \cite{Zhu2014, Fang2018}, and STM experiments on MoS$_2$ MLs \cite{Trainer2022}, interacts with the CB, by dynamically trapping and releasing electrons, if the frequency of the probing signal, $f_{RF}$ is lower than the cut-off frequency $f_C$. $f_C$ is the inverse of the defect trapping time $\tau=\frac{1}{f_c}=\tau_0 e^{{\Delta E}/{k_BT}}$. Here, $\tau_0$ is the trapping time constant, $\Delta E$ is the energy difference between the majority carrier band edge and the defect state, $k_B$ the Boltzmann constant, and $T$ the temperature \cite{Brammertz2008, Zhu2014}. Multiple defect states can occur at different $\Delta E$, each contributing to the carrier exchange on a different time scale \cite{Zhu2014}. For simplicity, the inset of figure \ref{fig:parabola}-bottom panel only shows a single defect state, henceforth called $A$. 

Thus, $c_S= c_Q + c_{SRH}$ (see inset of figure \ref{fig:parabola}-middle panel), where $c_{SRH}$ is the contribution of the defect states to the total sample capacitance. As derived by Lehovec \cite{Lehovec1966} within the Shockley-Read-Hall (SRH) theory, $c_{SRH}$ arises from the total admittance $Y$: $Y= c^*_{SRH}\frac{ln[1+(2\pi f_{RF})^2 \tau^2]}{\tau} + i 2 \pi f_{RF} c^*_{SRH}\frac{arctg(2\pi f_{RF}\tau)}{2\pi f_{RF}\tau}$ \cite{Lehovec1966}. The real part of $Y$ represents the defect states resistance, $R_{SRH}(f_{RF})=c^*_{SRH}\frac{ln[1+(2\pi f_{RF})^2 \tau^2]}{\tau}$, and the imaginary part corresponds to the defect states capacitance $c_{SRH}(f_{RF})=c^*_{SRH}\frac{arctg(2\pi f_{RF}\tau)}{2\pi f_{RF}\tau}$. Both $R_{SRH}(f_{RF})$ and $c_{SRH}(f_{RF})$ are frequency dependent. Here, $c^*_{SRH}=\frac{e^2 D_{SRH}}{2}$, with $D_{SRH}$ being the defect state density. 

Measurements of $c_S$ as a function of the gate voltage (equivalent to out $V_T$) in MoS$_2$ ML-based field effect transistors have shown that \cite{Zhu2014}: (i) defect states induce a symmetry breaking in $c_S$ vs gate voltage; (ii) $c_S$ can be approximated as a step function centered at zero gate voltage.

The middle panel of Figure \ref{fig:parabola} shows, in red, the plot of the $c_S(V_{DC})$ function used to model our experiments at $f_{RF}=0$: for $V_{DC}>V_0$, the DB acts as an electron reservoir, which screens the electric field effectively. In this regime, $c_S$ saturates to the degenerate value of $c_Q$, $c_{Q,deg}=e^2g_{2D} S =84\frac{\mu F}{cm^2}(\pi R_{tip}^2)$ \cite{Fang2018}. Here, $S= \pi R_{tip}^2$, with $R_{tip}$ being the tip curvature radius, is used as a reference surface area. Compared to $c_T$ (black dashed line in the middle panel of Figure \ref{fig:parabola}), $c_S(V_{DC}>V_0)=c_{Q,deg} >> c_T$, thus $V_T=V$ and the MoS$_2$ ML behaves like a metal, providing a full electric field screening.     

On the contrary, given the breaking of the electron-hole symmetry, due to the n-type nature of the MoS$_2$ ML, when $V_{DC}<V_0$, $c_S$ is much smaller: $c_S (V_{DC}<V_0)$ comes from electrons locally depopulating DB ($c_{SRH}$), with $c_Q=0$ until $V_{DC}$ is such that $V_S$ intercepts the valence band (VB) (this regime is beyond the scope of our work). As $c_S (V_{DC}<V_0)$ approaches $c_T$, making $V_T= \frac{c_S}{c_S+c_T}V$, the electric field screening is poor.

Here, we use $c_T(z)=2\pi \varepsilon R^*ln\left(\frac{z+R^*}{z}\right)$ , with $z=100$ nm and $R^*$ the effective radius (Supporting Information 3,4)\cite{Li2022}.

The breaking of electron-hole symmetry, characteristic of MoS$_2$ DOS, gives a $V_{DC}$-asymmetric $V_T$ (and, consequently, $V_S$), as shown in the inset of Figure \ref{fig:parabola}-top panel. At $V_{DC}>V_0$, $V_T$ matches $V_{DC}$ and $V_S=0$ (perfect electric field screening); at $V_{DC}<V_0$,  $V_T$ deviates from $V_{DC}$ ($V_T = \frac{c_S}{c_S+c_T} (V_{DC}-V_0)$), and $V_S=(V_{DC}-V_0)-V_T \neq 0$. Importantly, $eV_S(V_{DC}<V_0)< E_g$ in the $V_{DC}$ range we probe.  

The fit of the $df(V_{DC})$ spectrum of the MoS$_2$ bubble (red solid line in the top panel of Figure \ref{fig:parabola}), obtained by considering the asymmetric $V_T$, allows us to determine $c_S$. 

The asymmetry of $c_S(V_{DC})$ is further confirmed by the analysis of the cantilever oscillation amplitude $A$ (bottom panel of Figure \ref{fig:parabola}): $A_{Bubble}$ is reduced for negative $V_{DC}$ because of negative feed-back on the oscillation amplitude due to the small $c_S(V_{DC}<V_0)$. Details about the dynamical induced sample voltage and its effect on the cantilever oscillation as well as further insights into the asymmetric capacitive contribution of MoS$_2$ bubbles are reported in Supporting Information 5,6. Finally, we verified that the bubble height and diameter are independent of the applied bias voltages. 

A z-dependent analysis of the $df(V_{DC})$ spectroscopy of MoS$_2$ bubble, bulk, and Au is reported in Supporting Information 7. 

As  discussed above, the trapping and releasing of the electrons by the DB, as function of time, leads to a frequency dependent $c_{SHR}(f_{RF})$ and $R_{SHR}(f_{RF})$ \cite{Lehovec1966}. This translates into a frequency dependent $V_T$:

\begin{equation} \label{VRF}
V_T(f_{RF})=\frac{1+i 2 \pi f_{RF}\left[c_S\left(f_{RF}\right)\right]R_{SRH}\left(f_{RF}\right)}{1+i 2\pi f_{RF}\left[c_S\left(f_{RF}\right)+c_T\right]R_{SRH}\left(f_{RF}\right)}V 
\end{equation}

When an AC bias voltage $V_{RF}cos(2\pi f_{RF}t)$ is applied with $f_{RF}<f_c$, the DB will periodically fill and empty \cite{Brammertz2008, Izumi2023}. Conversely, for $f_{RF}>>f_c$, the contribution of the defects to $c_S$ is frozen out (i.e. $c_{SRH}$ goes to zero). Therefore, at sufficiently high probing frequency, the only contribution to $c_S$ is given by the quantum capacitance $c_Q$. This approach has been previously used by Izumi R. et al.\cite{Izumi2023} to measure the interface state capacitance (thus the interface state density) in a pn-patterned Si. By extracting $c_S$ from the $df(V_DC)$ spectroscopy, in the two frequency limits $f_{RF}<<f_c$ and $f_{RF}>>f_c$, and subtracting one from the other, the authors were able to decouple the bulk depletion contribution $c_D$ from the interface contribution $c_{int}$ to the total sample capacitance.

In the case of 2D materials, $c_Q$ and $c_{SRH}$ play the role of $c_D$ and $c_{int}$\cite{Izumi2023}, respectively, so that, in the low frequency limit, Eq. \ref{VRF} reproduces the DC case, $V_{T,0}=\frac{c_S}{c_T+c_S}V$ (with $c_S = c_Q + c_{SRH}$) while in the limits of high frequency, $V_{T,\infty}=\frac{c_Q}{c_T+c_Q}V$. These equations are formally the same as derived by Izumi R. et al.\cite{Izumi2023}. 

As the electrostatic force is non-linear in $V$, the response to a high frequency bias can be measured at low frequency. This alows us to access $c_Q$ and get rid of $c_{SRH}$. 

We studied the frequency-dependent electrostatic response of bubbles to an RF bias, $V_{RF}cos(2\pi f_{RF}t)$, with $f_{RF} = 0.1 \div 300$ MHz, by monitoring the time averaged cantilever frequency shift $\left\langle df \right\rangle = -\frac{f_0}{4k} \frac{\partial^2}{\partial z^2} (c_T \left\langle V_T^2 \right\rangle)$. Given $V=(V_{DC}-V_0)+V_{RF}cos(2\pi f_{RF}t)$, in the limit of  $V_{DC}=V_0$, and $\left\langle {\left[V_{RF}cos(2\pi f_{RF}t)\right]}^2\right\rangle =\frac{V_{RF}^2}{2}$: i) if the electric field is perfectly screened $V_T^2= V^2=V_{RF}^2/2$ and $\left\langle df \right\rangle$ is $f_{RF}$-independent; ii) if the screening of the electric field is not perfect, $V_T$ is $f_{RF}$-dependent, and so is $\left\langle df \right\rangle$. Here we assume that for $V_{DC} = 0$ and sufficiently small $V_{RF}$ , $c_S$ is constant.

\begin{figure}
    \centering
    \includegraphics[width=16cm]{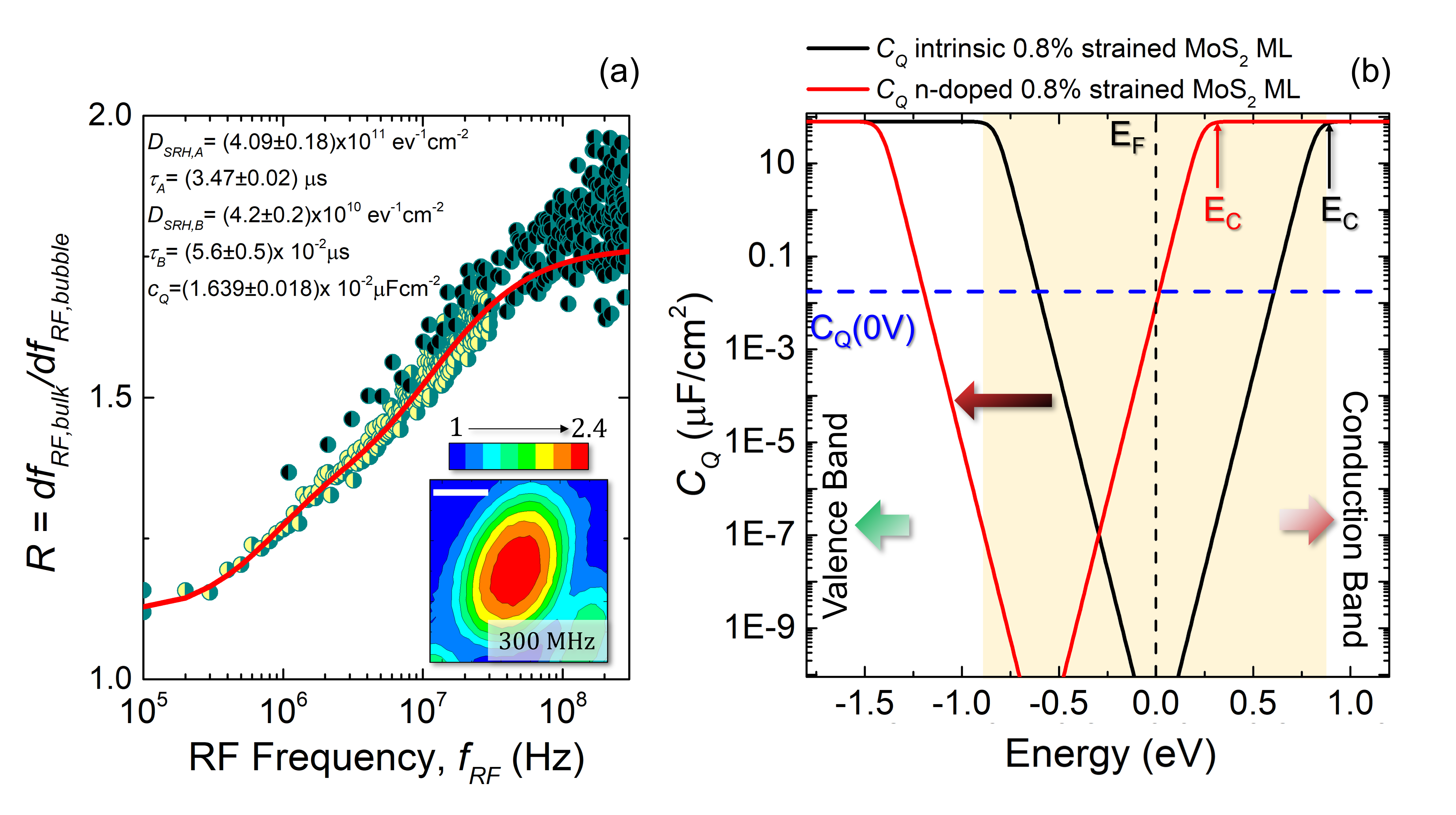}
    \caption{(a) Main: frequency shift ratio $R=\frac{df_{RF,bulk}}{df_{RF,bubble}}$ fitted, in red, to the ratio between Eq. (S.16a) and (S.16b). Different inner-scatter colors refer to two sets of data taken with smaller steps on a smaller frequency range (yellow), and with larger step in a larger frequency range (black). Fitting parameters and their standard error, evaluated as the statistical uncertainty from the fit, are reported. Inset: map of $R(x,y)$ at $f_{RF}$=300 MHz. Scale bar is $\mathrm{1 \mu m}$. (b) Black: theoretical plot of $c_Q(E)$ for an intrinsic semiconductor having $E_g$=1.65 eV, at $T$=300 K. Red: theoretical plot of $c_Q(E)$ for an n-doped semiconductor having $E_g$=1.65 eV, and $E_C-E_F \simeq$ 204 meV. Blue: experimentally measured value of $c_Q(0\ V)$. CB and VB appear at positive and negative energy, respectively. The yellow region indicates $E_g$. The thick horizontal red arrow points at the shift between the two curves. Black and red $E_C$ indicate the minimum of CB for intrinsic and n-doped semiconductor.}
    \label{fig:dfratio}
\end{figure}

Figure \ref{fig:dfratio}(a) shows the ratio between the measured $\left\langle df_{RF,bulk} \right\rangle$ and $\left\langle df_{RF,bubble} \right\rangle$,  $R=\frac{\left\langle df_{RF,bulk} \right\rangle}{\left\langle df_{RF,bubble} \right\rangle}$ measured at the center of the bubble (see Supporting Information 8 for individual $\left\langle df_{RF,bulk} \right\rangle$ and $\left\langle df_{RF,bubble} \right\rangle$ traces). The fit, in red, is to their analytical expression (Supporting Information 8). The resulting fit parameters, $c_Q$, $c_{SRH}(D_{SRH},\tau)$ and $R_{SRH}(D_{SRH},\tau)$, are shown in the figure. A good fit is achieved when assuming two defect states (A and B) at two different energies in the bandgap, and with different emission times $\tau$, (3.47 $\pm$ 0.02) $\mathrm{\mu s}$ and (5.6 $\pm$ 0.5) $\times 10^{-2}$ $\mathrm{\mu s}$, respectively. The fit to a single defect state is shown in Supporting Information 9. Thus, both lie in the MHz and sub-MHz range, where they dominate the electrostatic response of the material. 

For $f_{RF}>>f_c$, $V_{T,\infty} \simeq \frac{c_Q}{c_T+c_Q}V$, thus $R$ is mostly governed by $c_Q$. The black curve in figure \ref{fig:dfratio}(b) is the theoretical $c_Q(E)$ for a 0.8\% strained un-doped MoS$_2$ ML, which has $E_g=1.65$  eV \cite{Lopez2016}. Here $E_F \equiv 0$ eV, $E_g$ is the width of the shaded yellow region and $E_c$, the bottom of CB (black arrow). Thermal broadening at the band edges is also observable, due to the input temperature of $T=300$ K. Overlaid on the simulated quantum capacitance, the $c_Q$ value measured in our experiment at $V_{DC}=0$ V, and extracted from the fit of $R$, is plotted as the horizontal blue dashed line ($c_Q(0\ V) = (1.639 \pm 0.018) \times 10^{-2}\ \frac{\mu F}{cm^2}$). As no states are expected at the Fermi level in intrinsic semiconductors, the blue and black curves do not intercept each other at zero energy. However, for n-doped semiconductors the Fermi level is not in the middle of $E_g$, but closer the edge of CB. The red curve shows $c_Q(E)$, with the bottom of CB shifted toward $E_F$. It thus models the quantum capacitance expected in an n-type semiconductor, with $E_C -E_F = 204 $ meV. 

\begin{figure}
    \centering
    \includegraphics[width=16cm]{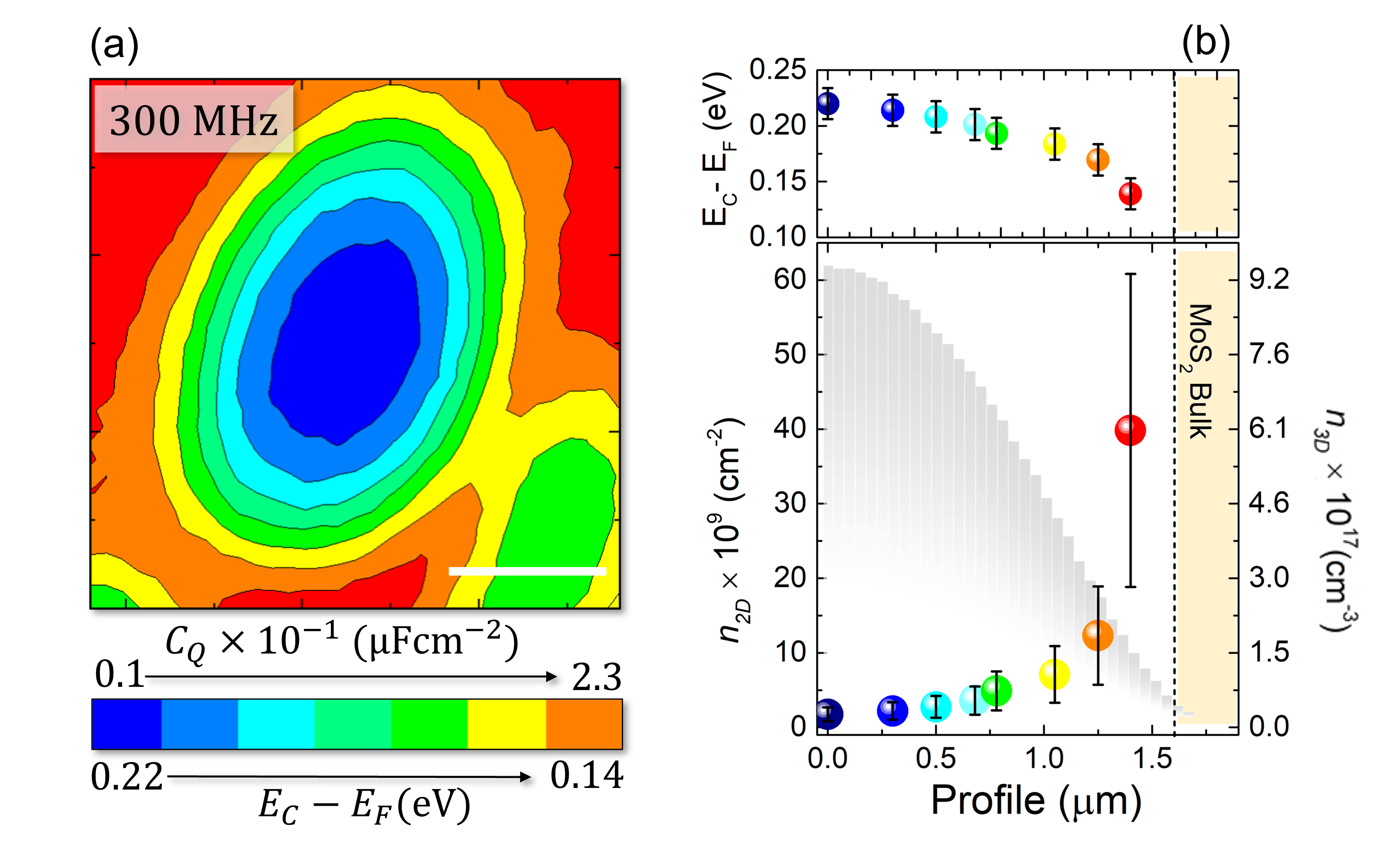}
    \caption{(a) Map of $c_Q(0\ V)$ obtained from the inset of figure \ref{fig:dfratio}(a) by using the ratio between Eq. (S.16a) and (S.16b). The evolution of $c_Q(0\ V)$ can be converted into the evolution of $E_c - E_F$. Note that in the red region (MoS$_2$ bulk) $c_Q$ and thus $E_c- E_F$ are not defined. Scale bar is 1 $\mathrm{\mu m}$. (b) Evolution of $E_C - E_F$ (top), 2D and 3D carrier density, along the bubble profile (bottom). The gray region in the background shows the bubble profile ($h_{max}$ =127 nm). The vertical dashed line indicates the transition from ML to bulk (in yellow). Errors bars have been calculated by propagating the error on $c_Q$, and assuming $E_g \pm 4\% E_g$ - as a consequence of a strain uncertainty of $\%\epsilon \pm 0.1$ - and $T \pm 1\% T$.}
    \label{fig:quantumcap}
\end{figure}

We determined $c_Q$ over the surface of a bubble by mapping $R(x,y)$ at $f_{RF}=300$ MHz (inset of figure \ref{fig:dfratio}(a)). At this frequency, the contribution of the defect states is negligible. $R$ evolves from a maximum of 2.4, at the center of the bubble, to 1, in the bulk region. We convert $R(x,y)$ to $c_Q(x,y)$ (figure \ref{fig:quantumcap}(a)), by using Eq. (S.16b) (at fixed $f_{RF}$, $c_{SRH}$ and $R_{SRH}$). The quantum capacitance is minimal in the center of the bubble and increases monotonically towards the edge, from $0.1$ to $\mathrm{2.3\times 10^{-1}\ \frac{\mu F}{cm^2}}$. 
The evolution of $c_Q$ from the center of the bubble (blue) to the edge (red) obtained from $E_g$ (extracted from Ref. \cite{Lopez2016}, see Supporting Information 1) is shown in Supporting Information 10. 
We then convert $c_Q(0\ V)$ to $E_c - E_F$ (color bar of figure \ref{fig:quantumcap}(a)). 

The variation of $E_c-E_F$ across the bubble is shown in top panel of figure \ref{fig:quantumcap}(b)). As $E_F$ is pinned, the behavior of $E_c-E_F$ reflects that of the CB minimum. Compared to the representative bubble profile shown in the background of figure \ref{fig:quantumcap}(b)-bottom, $E_C$ increases by $\sim$80 meV, between the bubble's edge and its center. Our result contradicts the expectations based on which $E_C$ should move close to $E_F$ in pristine MoS$_2$ MLs subjected to strain \cite{Hsu2017}. In our case, the reduction of $E_c - E_F$ while crossing the bubble must indicates an increase in n-type doping - and in the carrier density leaking from the sides - as we move from the center of the bubble towards the edge, hence as we release the strain. We believe that the observed behavior is governed by the role of the defects. For instance, it is shown that passivation of sulfur vacancies via hydrogen incorporation in MoS$_2$ MLs moves the top of VB toward $E_F$ \cite{Pierucci2017}. This could explain $E_{c,bubble}>E_{c,bulk}$. However, the evolution of $E_{c,bubble}$ from the center (maximal) to the edge (minimal) would indicate a locally higher H passivation activity at the center. Note that the preferential H-absorption at the bubble center (where strain is highest) agrees with the work of Li et al. \cite{Li2016}, which predicts a decreasing H-adsorption free energy with increasing tensile strain. 

Once $E_F$ is determined, one can calculate the 2D carrier density, $n_{2D}=g_{2D}k_BT ln\left[1+e^{{(E_F-E_C)}/{k_BT}}\right]$ \cite{Bera2019, Ma2015} and convert it into the 3D carrier density, $n_{3D}$, by dividing by the thickness of the ML (0.65 nm). The bottom panel of figure \ref{fig:quantumcap}(b) shows the evolution of 2D and 3D carrier densities across the bubble, whose height profile is shown in gray in the background. They increase from 1.7 (2.7) to 40 $\mathrm{\times 10^9\ cm^{-2}}$ (60 $\mathrm{\times10^{16}\ cm^{-3}}$), when moving from the bubble center to its edge. Here, the $n_{3D}$ we find is, within the experimental error, of the same order of magnitude of the one found on MoS$_2$ bulk via four-probe resistance measurements, and is confirmed by the EFM spectroscopy (see Experimental, figure \ref{fig:simulation}). Consistent with a higher CB edge, compared to $E_F$, the carrier density measured on the core of the bubble ($n_{3D}\sim 10^{16}\ cm^{-3}$) is lower than that in the bulk, in agreement with local passivation \cite{Pierucci2017}.

\section{Conclusions}

We used RF-EFM to investigate the electric field response of biaxially strained MoS$_2$ MLs in the form of bubbles. These bubbles were produced via H-ion irradiation of bulk crystals. The irradiation method induces strain in thick TMD crystals without the need to first isolate MLs through exfoliation or CVD growth. These latter are known to induce defect states in an uncontrollable manner, causing diffused dangling bonds and atomic scale defects (vacancies or grain boundaries). In contrast, low energy treatments performed in H-plasma have been demonstrated to desulphurize only the topmost interface of the ML \cite{Lee2021} and possibly passivate sulfur vacancies, due to H intercalation \cite{Pierucci2017}. Importantly, defect states affect the electrostatic response of MLs as much as their intrinsic quantum capacitance. Both contributions are DOS-related, and give insight into the material's band structure. By using RF-EFM we separated the contribution of defect states from that of quantum capacitance: the former dominates the electrostatic response in the sub-MHz range. Our data are compatible with the existence of defect states at two different energies in the bandgap, with characteristic emission times of 5.6 $\times 10^{-2}$ and 3.5 $\mathrm{\mu s}$, respectively. At higher frequency, the quantum capacitance dominates instead. We imaged $c_Q$ across a bubble surface, at $f_{RF}=300$ MHz, demonstrating the correlation between quantum capacitance (electrostatic properties, band structure) and bubble topography (strain). From the local value of $c_Q$ we obtained $E_C-E_F$, and showed that Fermi level and CB minimum move progressively closer together, when going from the bubble center towards the edge. Finally, from $E_C-E_F$ we obtained the charge carrier density ($n_{3D}$ or $n_{2D}$). This increases monotonically from the center to the edge of the bubble. This behavior is consistent with a possible local S-vacancy passivation, with a greater H passivation at the bubble center, where the ML is subjected to the maximum strain, in agreement with theoretical calculations \cite{Li2016}: the higher the tensile strain, the lower the H adsorption free energy of a MoS$_2$ strained surface. Note that, as shown in Ref. \cite{Bennet2023}, the variation of carrier density can severely limit the on-state capacitance of a MoS$_2$ ML-based device, to values well below its degenerate capacitance, and thus deserves attention for practical applications. 

In conclusion, imaging the quantum capacitance at finite frequency, through RF-EFM, allowed us to gain insight into the effect of H-irradiation on the local MoS$_2$ ML doping level and defect states, and its variation on a sub-micrometric scale. This technique can be useful to investigate diverse electrostatic phenomena at the mesoscopic scale over and up to microwave frequencies. Additionally, being nondestructive as a non-contact AFM technique, RF-EFM is robust and can be used for a wide range of applications.

\section{Experimental}

\subsection{Sample Preparation and Proton Irradiation}

The first superficial layers of commercial MoS$_2$ bulk crystals, from 2D Semiconductors, were first mechanically exfoliated with tape. Once a fresh and flat surface was obtained, the bulk crystals were placed in a vacuum chamber, equipped with a Kaufman ion source. The proton irradiation was performed at a base pressure of $<1 \times 10^{-6}$ mbar, and at a temperature of 120-150 $^\circ$C, with the crystal being grounded along the entire procedure. Hydrogen ions were obtained in an ionization chamber and accelerated by a system of biased grids, thus irradiating the sample with an ion beam having energy in the range of 10--20 eV. The samples were irradiated with a total dose in the range of 6-7 $\times 10^{16} ions/cm^{2}$. As reported in Ref. \cite{Tedeschi2019}, this procedure allows to form bubbles in TMDs with ML thickness.

\subsection{RF assisted-Electrostatic Force Microscopy}

The EFM measurements were performed in the air by using an attoAFM I, from Attocube, equipped with an interferometric sensor for the detection of cantilever deflection and/or oscillation. The data acquisition system is controlled by Nanonis electronics from Specs, interfaced with ACC100 and ANC350 modules, from Attocube, for powering the interferometric cavity, and driving step motors and piezo movements. 

We used conductive solid platinum AFM probes from Bruker (RMN-25PT400B), having resonance frequency in the range $f_0=5 \div 11$ kHz, quality factor of $Q=255$, and spring constant in the range $k=6 \div 14$  N/m. With the tip holder solidly connected to an excitation piezo, we preliminarily characterized the cantilever resonance curve by monitoring the cantilever oscillation amplitude, throughout the interferometric sensor, while sweeping the frequency of the AC bias supplied to the piezoelectric actuator. Upon identifying the cantilever resonance frequency, we worked in a closed loop mode: phase locked loop (PLL) control circuit was used to keep the phase constant; the cantilever oscillation amplitude was kept constant ($\sim$10 nm, much smaller than the tip-surface distance) by adjusting, if needed, the amplitude of the AC bias supplied to the piezo-actuator (figure 1d). 

\begin{figure}
    \centering
    \includegraphics[width=10cm]{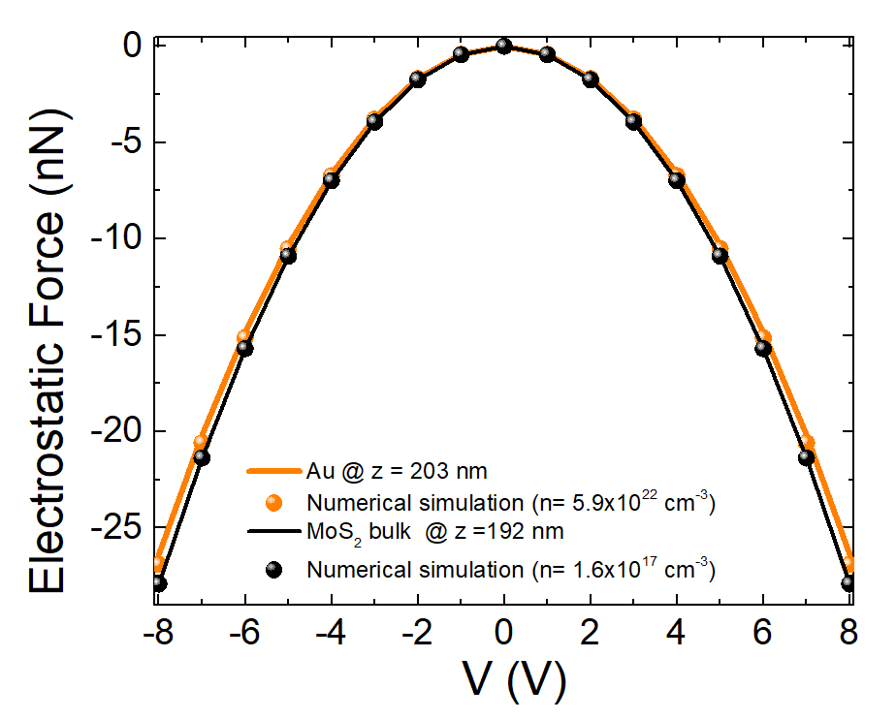}
    \caption{$F_{el}(V)$ spectroscopy (solid lines) and numerical simulations (scatters) of tip-Au and tip-MoS$_2$ electrostatic interaction, at 203 and 192 nm, respectively, obtained by using the values of charge carrier densities $n$ reported in the legend, and an effective tip radius of $3.02 \pm 0.01 \mu m$ - see Supporting Information 3}.
    \label{fig:simulation}
\end{figure}

We measured the tip-sample electrostatic interaction, and its spatial and bias dependence, throughout the variation of the frequency of the oscillating cantilever, $df=f-f_0$, with $f_0$ and $f$ being the free resonance frequency and its interaction-driven shifted value. The frequency shift $df$ is indeed related to the interaction force, electrostatic in our case, $F_{el}$, as $df=-\frac{f_0}{2k}\frac{\partial F_{el}}{\partial z}$. The topographic measurements were acquired in closed-loop mode, i.e., by keeping, pixel by pixel, a constant $df$ (constant tip-surface separation). To do so, we first set up a tip-surface interaction satisfying the condition of negative $df$ (attractive interaction), at fixed supplied bias $V_{DC}$ ($F_{el}\propto V_{DC}^2$). We chose $df=-10\ Hz$ and $V_{DC}=2\ V$, which guarantees a tip-surface distance of $\sim$100 nm (well outside the vdW force range). Subsequently, a feedback loop was used to counterbalance for the variation of tip-surface distance, occurring while scanning due to the topographic features. The feedback circuit biased the scanning piezo so as to induce displacements along the z-direction counteracting for the morphological variations, and thus preserving the strength of the electrostatic interaction and the initial value of the frequency shift $df$. Conversely, in spectroscopic mode, the feedback loop was disabled and the variations of $df$ (vs $V_{DC}$ or $f_{RF}$) were recorded at fixed tip-surface distance. We used the measured $df(V_{DC})$ spectra of MoS$_2$ bulk and Au, at tip-sample separations of 192 and 203 nm, respectively, and their fit to quantify $\frac{\partial^2 c_T}{\partial z^2}$. This latter was used to calculate the electrostatic force, $F_{el}$, acting between the AFM probe and both samples, at fixed separation, resulting in the solid lines of figure \ref{fig:simulation}. By describing the tip shape as an axially symmetric paraboloid, Xu et al. \cite{Xu2019} have numerically simulated the electrostatic force acting between a metallic AFM probe and a sample with known carrier density, $n$. Following their formalism, we have simulated $F_{el}$ by assuming a charge carrier density of $n_{Au}=5.9\times 10^{22}\ cm^{-3}$ \cite{AuDatabase} and $n_{MoS_2}=1.6\times 10^{17}\ cm^{-3}$ (the latter extracted from four-probe resistance measurement on a twin MoS$_2$ crystal). The numerical simulations (scatters in figure \ref{fig:simulation}) reproduce the experimental measurements (solid lines), thus demonstrating that EFM experiments can also lead to the quantitative evaluation of the charge carrier density. 

Finally, the map of $R$(300 MHz) - figure \ref{fig:dfratio}(c)-left - was obtained by acquiring the value of $df$ (at $300$ MHz), per each pixel, at the same tip-sample separation. To do so, the initial feedback conditions ($df=-10$ Hz and $V_{DC}=3$ V) were always restored before moving from one pixel to the next, so as to ensure that the same tip-surface distance ($\sim$153 nm) is set up per each point. Once moved, the feedback was disabled, the bias was swept to the desired conditions, and the local value of $df$ was measured (at fixed height).

 We used arbitrary wave function generators (AFG3252 from Tektronix or Hameg HM8134) to supply the tip with a sinusoidal voltage signal, by keeping the crystal grounded. The output signal was first amplified (through Mini-Circuits Amplifier ZLF-2500 VH+) and then summed up to the DC bias supplied by Nanonis, throughout the use of a bias tee (ZFBT-6GW from Mini-Circuits - inset of figure 1b). Finally, it travelled to the tip within a microwave coaxial cable. As discussed in Supporting Information 8, we found a RF-frequency dependent attenuation of the RF power, which decreases as $f_{RF}$ increases (as also shown in the top panel of figure 4(a)). We also found that, differently from the case of MoS$_2$ bubble, the $df(f_{RF})$ spectra measured on Au and MoS$_2$ bulk crystals are almost undistinguishable, and thus only affected by the attenuation of the RF line.

Finally, as discussed in Supporting Information 3, we have adopted a sphere-plane (sphere-sphere) model to describe the electrostatic interaction between the tip and Au or MoS$_2$ bulk crystal (MoS$_2$ bubble). By doing so, we have analytically modelled the quantities involved in the problem, such as $F_{el}$ and $c_T$ (and its derivatives).

\begin{acknowledgement}

C.D.G. acknowledges financial support from MUR (Italian Ministery for University and Research) under the project PON-AIM (Programma Operativo Nazionale - Attraction and International mobility). M.A. and C.Q.H.L. acknowledge financial support from the French Agence Nationale de la Recherche through the ANR JCJC grant (SPINOES). 

\end{acknowledgement}

\begin{suppinfo} 
\setlength{\parindent}{0cm}
\textbf{1. Bubble characterization}: Topography and strain analysis of MoS$_2$ bubble.

\textbf{2. Mapping the contact potential difference: Kelvin probe
force microscopy}: Comparison between topography and contact potential map, as obtained by two-pass KPFM.

\textbf{3. Analytical description of tip-sample electrostatic interaction}: Tip-sample electrostatic force and cantilever frequency shift derivation.

\textbf{4. Capacitance comparison}: Quantitative comparison among the capacitance terms involved in the RF-EFM experiment.

\textbf{5. Derivation of bias-sign dependent tip-MoS$_2$ bubble electrostatic
interaction}: Derivation of tip-bubble electrostatic interaction at $V_{DC}>0$ and $V_{DC}<0$.

\textbf{6. Cantilever resonance curves modelling}: Derivation of cantilever dynamic as a harmonic oscillator.

\textbf{7. Tip-surface separation dependence of $df(V_{DC})$}: $df(V_{DC})$ spectroscopy of MoS$_2$ bubble, bulk and Au, as a function of tip-surface separation.

\textbf{8. Calibration of RF power}: Calibration of df vs RF voltage and of RF voltage at fixed RF frequency.

\textbf{9. Fit of $df$ ratio vs $f_{RF}$ spectroscopy - single vs double
defect states}: Validation of double defect state fit.

\textbf{10. Quantum Capacitance}: Plot of $c_Q(E)$ as a function of position along the bubble profile.

\end{suppinfo}

\bibliography{main}

\end{document}